# Mining Reddit Data to Elicit Students' Requirements During COVID-19 Pandemic

Shadikur Rahman
*EXINES Lab*
York University
sadicse@yorku.ca

Faiz Ahmed
*EXINES Lab*
York University
faiz5689@my.yorku.ca

Maleknaz Nayebi
*EXINES Lab*
York University
mnayebi@yorku.ca

*Abstract*—Data-driven requirements engineering leverages the abundance of openly accessible and crowdsourced information on the web. By incorporating user feedback provided about a software product, such as reviews in mobile app stores, these approaches facilitate the identification of issues, bug fixes, and implementation of change requests. However, relying solely on user feedback about a software product limits the possibility of eliciting all requirements, as users may not always have a clear understanding of their exact needs from the software, despite their wealth of experience with the problem, event, or challenges they encounter and use the software to assist them. In this study, we propose a shift in requirements elicitation, focusing on gathering feedback related to the problem itself rather than relying solely on feedback about the software product. We conducted a case study on student requirements during the COVID-19 pandemic in a higher education institution. We gathered their communications from Reddit during the pandemic and employed multiple machine-learning and natural language processing techniques to identify requirement sentences. We achieved the F-score of 0.79 using Naive Bayes with TF-IDF when benchmarking multiple techniques. The results lead us to believe that mining requirements from communication about a problem are feasible. While we present the preliminary results, we envision a future where these requirements complement conventionally elicited requirements and help to close the requirements gap.

## I. INTRODUCTION

In the face of unknown events, software can play a crucial role in helping people adapt and find solutions. By leveraging the power of technology and data, the software can assist in various ways to address unforeseen challenges. This holds true for the COVID-19 pandemic, where the requirements of students were largely unknown and evolving rapidly. During the initial stages of the pandemic, educational institutions worldwide faced unprecedented disruptions. Schools and universities had to quickly transition from traditional classroom settings to remote learning environments. Software played a vital role in facilitating this shift [5]. Additionally, software applications and platforms were developed to address the specific needs of students during the pandemic [11]. Furthermore, data-driven software solutions were instrumental in monitoring and managing the spread of COVID-19 within educational institutions. Contact tracing apps and health monitoring systems helped identify potential virus exposure, allowing for timely intervention and prevention of further transmission [21]. All of these designs occurred through the rushed development of task forces and groups in educational institutes, where stakeholders attempted to ascertain the needs of students through assumptions and an error and trial process [20].

Motivated by the vision of data-driven requirements elicitation [23], [28], we explored the potential of utilizing social media as a means to elicit requirements in this particular context. Data-driven requirements engineering relies on the gathering and elicitation of requirements from crowdsourced data *related to software products*, such as those found on app stores [16], [30], Twitter [40], Reddit [13], or open forums [2], [29], [32]. This data specifically pertains to a software product of interest. For instance, Nayebi et al. [26], [27] conducted a study where they extracted requirements, including the number of apps, for the "Pokemon Go" app from both Google Play and Twitter. Khan et al. [19], [18] explored approaches using NLP for requirements discovery and classification. They introduced a crowd-based requirements engineering approach called CrowdRE-VArg. For that, they used unstructured text from Reddit for a specific Google Map mobile app. In their work, they identified software requirements for a specific domain. Hence, we explored the methods for shifting the focus of requirements elicitation from *information about software* to *information about an event or problem*, here, the COVID-19 pandemic, which requires a software solution. We propose this shift by recognizing that users may not have a precise understanding of how and when software can assist them (the solution), but they know well about their experienced problems. As a result, gathering input solely related to software (such as feedback, surveys, interviews, documents, etc.) does not result in a comprehensive set of elicited requirements. Instead, we propose using general-purpose communications about a topic, event, or problem that is unrelated to software as the source for the elicitation task.

However, this task is rather challenging. Compared to user feedback about a software product, general-purpose communications discussing an event such as the COVID-19 pandemic are highly noisy and include irrelevant information to the user's need. When eliciting requirements from mobile app reviews [8], [36] or a thread in the open forums [7], the discussion which sources the elicitation task is highly concentrated around the topic of a software product. Even then, the fake, spam, uninformative, and irrelevant comments reduce the performance of our automated methods [24]. Traditional Goal-driven requirements engineering methods such as KAOS [38] (Knowledge Acquisition in Automated Specifications) and NFR Framework (Non-Functional Requirements Framework) emphasized the importance of eliciting requirements by aligning them with business goals [6]. However, in recent times, this perspective seems to have been overshadowed by the data-driven approach to mining repositories for requirements elicitation, which only focuses on mining feedback and open forum discussions.

In this paper, we benchmark with models to mine requirements from unstructured communications. If we are successful in this task, we hope to further apply this approach for the automated recommendation of software services. In particular, we investigate two research questions:

**RQ1:** Which model performs the best for identifying student's requirements from Reddit?
**Why and How:** The identification of requirements sentences from a Reddit post is a binary classification task. Each sentence might communicate a need or a requirement (requirement sentence) or not (non-requirement sentence). Hence, to answer this question, we designed a case study on Reddit as a forum that evolves and shapes around emerging events [31]. We then manually labeled 3,118 sentences as requirements or non-requirements. We performed a comprehensive benchmark using different text embedding models. We experimented with the stateoftheart models and implemented five different classifiers to identify the requirements and compared Logistic Regression, Multinomial Naive Bayes, SVM, Random Forest, and K-Neighbors to identify the best model.

**RQ2:** What features can best model user requirements within Reddit?
**Why and How:** Feature selection is one of the most important tasks for machine learning problems as it improves the performance of the machine learning models and enhances the ability to predict training models by selecting the most important features. We performed a comprehensive benchmark using different feature selection models TF-IDF, Word2vec, Sentence-BERT, and USE to extract the most important features from unstructured communications. On the other hand, we selected the interrogative sentences (III-D) and requirements keyword-based sentences (III-C) as features for identifying requirements.

We do not believe that requirements mining replaces traditional requirements as described above but rather serves as a complementary process to understand users' needs more. Our research will contribute to improving requirements engineering by addressing the challenge of extracting requirements from unstructured communication in a comprehensive and organized manner.

In what follows, we first discuss the related work, then in Section III, we detail the design of our proof of concept case study. We follow in Section IV with other methodology for answering the research question and present our results in Section V. Lastly, we discuss the limitations of our study (Section VI) and explore the likelihood of our vision coming to fruition (Section VII).

## II. RELATED WORK

In this section, we present a summary of existing studies on techniques used in crowdsourcing requirements engineering, requirements classification, and natural language document classification.

### A. Crowdsourcing requirements engineering

Crowdsourcing has emerged as a promising approach to requirements engineering, allowing for the collaboration and collective intelligence of a large group of individuals. Several studies have explored the application of crowdsourcing at various stages of the requirements engineering process. Dalpiaz et al. [9] proposed a method called RE-SWOT to extract requirements from app store reviews by analyzing competing products. RE-SWOT combines natural language processing (NLP) algorithms with data visualization techniques. The tool was evaluated with expert product managers from three mobile app companies, and preliminary results indicate that competitor analysis is a promising method that has a direct impact on requirements engineering for app development companies. This research highlights the potential of using NLP and competitor analysis to make real-time decisions based on user feedback and improve app development processes. Rahman et al. [33] introduce the Example Driven Review Explanation (EDRE) method to enhance the code review process by providing additional explanations through examples. The results demonstrate that EDRE can assist software teams in improving code review quality in GitHub projects and reducing communication overhead to identify code issues. Kaliamvakou et al. [17] examined the use of crowdsourcing for requirements elicitation. They introduce the concept of "CrowdRE" and conduct an experiment to evaluate the quality of requirements gathered through crowdsourcing compared to traditional methods. These results indicated that crowdsourcing achieved comparable or better quality requirements, demonstrating its potential in the elicitation phase. Cerulo et al. [4] illustrates the semi-automated forum management (AFM) system, designed to establish discussion threads in open forums, exploring it through a series of case studies and experimental evaluations using feature requests quarried from open source forums.Li et al. [22] presents a classification method to convert user requests collected from crowdsourcing platforms (sourceforge.net) into structured software requirements. The proposed method uses project-specific and non-project-specific keywords with machine learning algorithms to achieve high accuracy, reduce manual effort, and manage minority classes effectively. By considering and building on these previous works, our approach aims to contribute to the field of crowdsourcing for requirements engineering by leveraging SubReddit data and machine learning techniques to extract valuable user requirements during the COVID-19 pandemic.

## B. Mining Reddit Data

In recent years, Reddit has emerged as a promising avenue for data mining research, providing a rich source of information in various domains, including software-related studies and non-software-related investigations. Khan et al. [19], [1], [18] introduced a crowd-based requirements engineering approach called CrowdRE-VArg, which uses user-generated information from Reddit forums to identify conflict-free new features, design options, and issues related to software applications. The method uses argument theory to analyze discourse and gradually assess the strength of supporting and attacking arguments. The proposed method helps resolve conflicts over new features or issues by finding settlements that satisfy crowd-user stakeholders. The research introduces an extended abstract argumentation and valuation paradigm called bipolar gradual valuation argumentation. A sample discussion on the Reddit topic concerning the Google Map mobile app is used to demonstrate the proof-of-concept. Melton et al. [25] study analyzed COVID-19 vaccine discussions on Reddit, finding predominantly positive sentiment and a focus on side effects rather than conspiracy theories. The results highlight the need for targeted messaging and interventions to address vaccine hesitancy. Jeong et al. [15] introduces an opportunity mining approach from for product planning using topic modeling and sentiment analysis of Twitter and Reddit data. It aims to systematically identify latent product topics, evaluate customer satisfaction, and identify product opportunities based on customer needs, contributing to real-time monitoring and analysis of evolving product environments. Drawing inspiration from these studies, our research explores the potential of mining Reddit data to elicit student requirements during the COVID-19 pandemic. Using machine learning techniques and analyzing user data, we aim to extract valuable insights that can contribute to understanding student needs.

## C. Natural language processing in requirements engineering

The use of natural language processing for eliciting requirements is widely popular. In particular, Janssens et al. [14] conducted a comprehensive literature analysis on the application of natural language processing (NLP) in requirements elicitation and analysis. The research and practice of requirements engineering have dynamically sought methods for requirements engineering, which have increasingly evolved with the emergence of data-driven methods. Zhao et al. [43] and [39] articles are presented in a mapping study that surveys the landscape of natural language processing for Requirements Engineering research (NLP4RE) for an overall understanding of the field. The results of this Mapping study can benefit researchers and practitioners in many ways. Ferrari et al. performed a systematic study on the use of NLP in requirements engineering [10]. Zamani et al. [41] conducted a mapping study of empirical studies to determine the effectiveness of machine learning techniques in improving the requirements of engineering processes and artifacts. Zhao et al. [42] provided the classification tasks and linguistic analysis of the natural language processing (NLP) technique based on their 57 NLP techniques. These systematic literature reviews as well as our independent search, shows that elicitation methods consistently evolved around a better understanding of users through mining their feedback about software (e.g. in mobile app stores [23] or open forums [7]) or better techniques for interviewing, surveying, or brainstorming with current or prospective software users. By acknowledging the existing research in NLP for requirements engineering, we demonstrate how our approach builds upon and contributes to the ongoing advancements in the field.

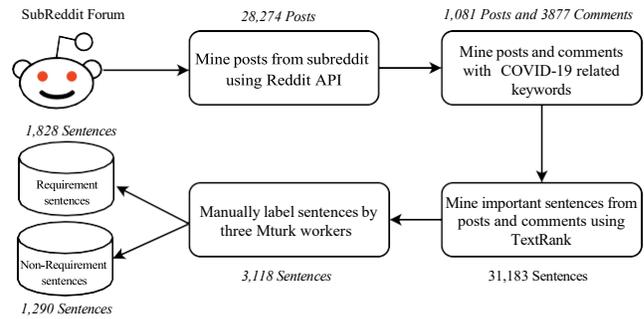

Fig. 1. The process flow of data gathering and labeling of our research.

> *While mining forums, social media, and in general crowd-sourced information are well established for requirements elicitation, they all are focused on communication about software (for instance, a mobile application).*

## III. Empirical Design and Data

We chose to design our initial case study on Reddit communication around the COVID-19 topic. In particular, Reddit allows for the creation of forums on user-selected topics, also known as SubReddits. The study conducted by Priya et al. [31] reported a significant variation between Twitter and Reddit regarding user needs. Their findings highlighted Reddit as a suitable platform for gaining comprehensive insight into multiple viewpoints within a limited time frame. The subReddit structure of Reddit is customized to cater to a particular interest.

Iqbal et al. [13] conducted an exploratory study on Reddit data about software products by mining and retrieving posts, including the name of 20 software products. They manually classified 1,915 posts into feature-related, bug-report, or irrelevant categories. They found that 54% of posts contained informative content about a bug report or a feature request. Unlike them, we are not interested in software-related posts. Specifically, for our case study focused on mining Subreddit posts and comments about events, we chose to utilize Reddit as our data source.

We chose this platform as a proof of concept for our experiment and focused on a higher education institute subReddit to provide additional context and coherence to the communication we analyzed. As a timely subject, we chose to design our proof of concept case study on the mining the students need in our affiliated institute (York University[1]). Students and faculty are very active on York Institute's subReddit. During the COVID-19 pandemic, announcements and communications have been distributed through this channel unofficially.

Our goal was to gather all the communications from the subReddit of York Institute during COVID-19 and identify the ones relevant to the pandemic and the circumstances caused by it. Then we use those Reddit posts to identify requirements and answer our research questions. Figure 1 shows the process flow of data gathering and labeling of our research.

*A. Data gathering process*

We used web scrapers to gather data from the York University subReddit forum between March 1st, 2020, and September 1st, 2021. This time frame covered the period of the institute's COVID-19-related closure and subsequent announcement of the return to in-person classes. To perform the data extraction, we relied on `Python` programming language and a variety of libraries, including `Pandas` and `BeautifulSoup`. In particular, we used `psaw`[2] and `praw`[3] APIs to mine posts from the subReddit of York institute. Both `psaw` (Python Pushshift.io API Wrapper) and `praw` provide `Python` API wrappers for Reddit and are commonly used to extract data from the Reddit forum. We combined these APIs to overcome post-fetching limitations. As a result, we gathered an overall of 28,274 posts from our targeted Reddit subforum. In Reddit, the initial submission on a new thread of communication is considered a "post", and follow-ups are known as "comments".

Not all these communications are relevant to the context of the pandemic. For instance, students often ask if they should take a particular course within a specific semester or faculty. Hence, we formed a set of keywords to mine relevant to the COVID-19 pandemic and specifically gathered posts and comments containing relevant COVID-19 keywords[4]. Then, we searched for the list of keywords commonly found on the internet for the COVID-19 pandemic. Inspired by that and our observations from Reddit, we built a list of 23 keywords that we searched within the collected Reddit posts and comments. These keywords were *Virus*, *COVID-19*, *COVID19*, *COVID*, *Pandemic*, *Corona*, *Symptoms*, *Flu*, *Screening*, *Exposure*, *Health*, *Accommodation*, *Online exam*, *Online test*, *Cases*, *Lockdown*, *Quarantine*, *Isolation*, *Infected*, *Outbreak*, *Medical*, *Antigen*, and *PCR*. Searching these keywords, we retrieved an overall 1,081 posts (initial post) and 3,877 (follow-up) comments from the subReddit forum, which contained at least one of these keywords. From here on, we refer to this set of relevant posts and comments as our dataset for the case study.

[1] https://www.yorku.ca/
[2] https://psaw.readthedocs.io/en/latest/
[3] https://praw.readthedocs.io/en/stable
[4] https://ritetag.com/best-hashtags-for-coronavirus

*B. Data labelling process*

To perform supervised classification, we labeled a subset of our Reddit data. Our research question involves a binary classification task to identify whether a sentence communicates a requirement or not. Hence, we parsed every Reddit post in our dataset into sentences using the sentence tokenizer of NLTK. The corpus contains small, unnecessary, and sometimes unfinished sentences. To ensure that we label high-quality data, we used three different text summarization algorithms SumBasic, TextRank, and LexRank. we evaluated our models performance based on ROUGE-1, ROUGE-2, ROUGE-3, and ROUGE-4. Then, we analyzed the performances of these three summarization techniques, TextRank algorithm has been determined as most suited to summarize our datasets. For that, we gathered the most important sentences of each post and comments using TextRank. As a result, we obtained a total of 31,183 sentences related to COVID-19 from 1,081 posts and 3,877 comments.

After that, we used `Amazon Mechanical Turk (MTurk)` to label our data. First, we randomly sampled 10% of all the sentences. As a result, we submitted 3,118 sentences to `Mturk` for manual labeling. Then, we designed a labeling task where a worker should decide if the sentence provided to them communicates a requirement or not. We first show each worker a sample set of sentences communicating a requirement (for example, *I don't know what root to take*, *How can I lose weight?*) and a sample that does not (for example, *did you walk your dog this morning?*, *I like my car.*). Then, we showed five sentences to each worker and asked them for each sentence to identify if the sentence was communicating a requirement or not. Figure 2 shows an example of the labeling task.

We submitted the 3,118 sentences for labeling by three independent workers. The final label was determined by a majority vote. Therefore, if two or more workers identified the sentence as a requirement (or not), we labeled the sentence accordingly. For example, "any chance a Covid-19 vaccine site will open on campus?" if two workers identified its requirement as "Yes" and another worker identified non-requirement as "No". In this case, we labeled it as a requirement sentence. If two or more workers selected "NO" for a sentence, then we labeled those sentences as non-requirement sentences. As a result, we labeled 1,828 sentences as a *requirement sentence* and 1,290 sentences as a *non-requirement sentence*.

*C. Sentences with the requirements keyword*

We also invested in identifying if specific keywords can lead us to identify requirements sentences. Keywords and term extraction for eliciting requirements [34] and later for tracing requirements [12] have been discussed in the field. Two annotators manually inspected the requirements sentences, each extracting a list of keywords associated with a requirement. For example, keywords such as *need*, *require*, *wish*, *problem*, *request*, *trouble*, *endeavor*, *interrogate*, *establish*, *inquire*, *refer*, *aim*, *remove*, *option* are highly relevant to the statement of requirements. By manual inspection, we identified 51 keywords. We then used `WordNet` to extend this list using the synonyms and relevant words to this core set. As a result, we build a set of 247 requirement-related keywords.

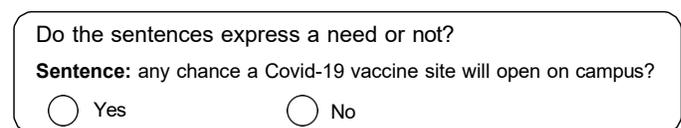

Fig. 2. A sample of labeling task assigned to the `Mturk` workers.

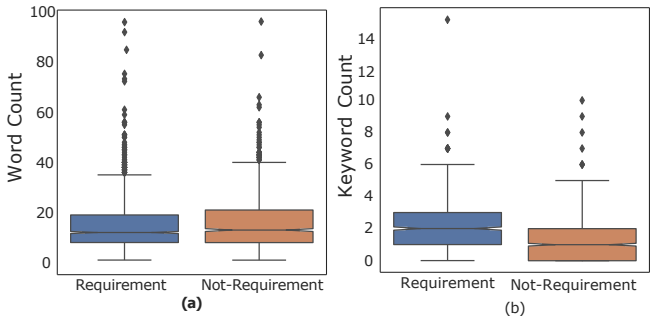

Fig. 3. Boxplot distribution of requirements keyword and word count in our labeled data

When looking into our labeled dataset of 3,118 sentences, we could identify 1,525 sentences with at least one of these requirement keywords. Among them, 895 sentences have been labeled as requirement sentences by our workers (58.6%). Figure 3 represents the overall requirements keyword count by requirement sentences. Figure 3 illustrates the distribution of the number of words and keywords in sentences, both for requirement and non-requirement sentences.

For measuring the significance between two variables(word count and keyword-based) with labels, we employed the Mann-Whitney U test. Our analysis revealed a significant difference between the two groups (requirement and non-requirement) in terms of the variables "word count" and "keyword-based," as indicated by a p-value of 0.049. This finding suggests that these variables play a substantial role in distinguishing between the two groups. It can be observed from Figure 3-(a) that non-requirement sentences are notably longer and contain more words. Additionally, Figure 3-(b) indicates that non-requirement sentences have significantly fewer requirement keywords. We further labeled each sentence with at least one keyword in it as "has a keyword" and otherwise as "does not have a keyword".

### D. Identifying interrogative sentences

The use of interrogative sentences in requirements engineering when analyzing user interviews has been discussed by Sultan and Miranskyy [37]. They explore the role of questions in eliciting requirements and highlight the potential of interrogative sentences in capturing user needs. Therefore, we hypothesized that often interrogative sentences communicate requirements or a need that is subject to a question.

For mining the interrogative sentences, we used the `Stanford CoreNLP` parser. Then, we selected SBARQ and SQ clause tags for classifying interrogative sentences. We considered these two clauses because the SBARQ clause tag is a direct question introduced by a wh-word or a wh-phrase, and the SQ clause tag is an Inverted yes/no question. We used these two clause tags for mining interrogative sentences to find our corpus. Apart from that, we retrieved question mark sentences from the unprocessed text; for example, sentences such as "How can I lose weight?" which end with a question mark. As a result, we obtained 237 sentences that have been classified as interrogative sentences and 2,881 sentences as they are "not a question". Among them, 160 interrogative sentences were labeled as requirements sentences. As a result, we also annotated each sentence as being an "interrogative sentence" or "not an interrogative sentence".

Figure 4 shows the overview of requirements sentences versus interrogative sentences (A ∩ B) and versus sentences with requirement keywords (A ∩ C).

In our study, we considered the sentences with requirement keywords (Section III-C) and interrogative sentences (Section III-D) as additional features for improving the performance of our model. For that reason, we focused on identifying key features through the analysis of requirement-based keyword sentences and interrogative sentences. We aim to extract valuable information about user needs and preferences. Keyword-based sentences allow us to capture clear references to desired features, providing a clear indication of user needs. on the other hand, Interrogative sentences helped us identify user inquiries and suggestions regarding potential areas of improvement. The use of these two features makes identifying requirements more efficient and effective.

## IV. EMPIRICAL METHODOLOGY

To answer our research question, we employed supervised learning using text embeddings as our method. In the following sections, we describe our approach and the additional features used to enhance the classifier's performance.

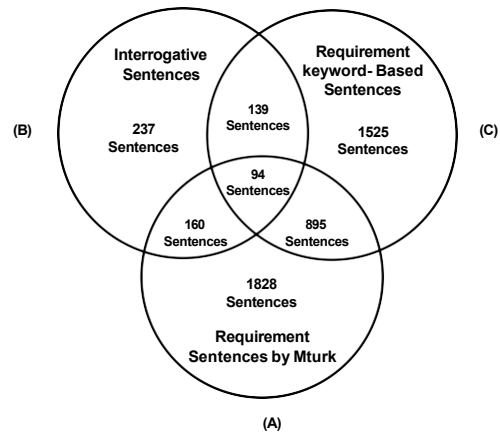

Fig. 4. Overview of the Requirements sentences (A) vs interrogative sentences (B) vs requirements keyword-based sentences (C)

## A. Supervised classification using textual features

We used natural language processing (NLP) techniques to identify sentences that state a requirement from general-purpose unstructured communication. To benchmark our approach, we evaluated the performance of five classification techniques and two feature selection models in this task. An overview of our methodology is presented in Figure 5. For these implementations, we used `Python`, `NLTK`, and `Scikit-Learn`.

*1) Text Prepossessing:* We applied a standard text preprocessing pipeline (Step ①) to our sentences, which included punctuation removal, typo mistake correction, tokenization, noise removal, POS tagging, lemmatization, and extraction of n-grams. We removed unnecessary punctuation from our sentences and used `autocorrect` python library for typo mistake correction from our sentences. Then, we applied the tokenization process to split words within each sentence. We also removed stop words from the text using common sets. We then used the lemmatization process to map words to their base or root form. Next, we generated a set of sequential word combinations from the input sentences using n-gram models. Specifically, we used bi-grams to maintain the sequence between two words, such as "How can", "can I", "I lose", and "lose weight" from the sample sentence "How can I lose weight".

*2) Feature extraction:* After preprocessing the text, we performed feature selection on the text in Step ② of our process to train our models. This step is necessary to convert the textual data into vectors that can be processed by the machine. In our research, we explored two types of feature extraction approaches, word level and sentence level, for identifying requirements. By considering different levels of embedding, we aimed to capture a more comprehensive understanding of the data and compared which embedding performed better for our training models.

*Word Embedding:* For word embedding, we experimented with both TF-IDF for weighting words and Word2Vec as an embedding model. Both of these models provide text representation for our classifiers. TF-IDF relies on the term frequency to extract important features. Word2Vec, however, creates word embeddings and adjusts weights based on forward and backward propagation and error gradients. TF-IDF works well when the text data is relatively small. On the other hand, Word2Vec is better at capturing the semantic relationships between words. We used both models to select features and train our classifiers.

*Sentence Embedding:* For Sentence Embedding, we used the Sentence-BERT and Universal Sentence Encoder (USE) to train our models. Sentence-BERT [35] is a sentence embedding technique that utilizes the BERT network. It uses a Siamese Network Architecture to enable the generation of high-quality sentence embedding. Universal Sentence Encoder (USE) is another sentence embedding technique by Google Research [3]. we examined both sentence embedding methods to train our classifiers.

*3) Sentence Classification Process:* We have a binary classification task in hand to determine if a given sentence is a requirement or not. We experimented with five classifiers for the supervised learning task, which were: Logistic Regression, Naive Bayes, Linear SVC, Random Forest, and k-nearest neighbor. To train our models, we created a balanced dataset using our labeled data as the ratio between non-requirement(1290 sentences) to the requirement(1828 sentences) was 0.70. To address the imbalance in our dataset, we used SMOTE(Synthetic Minority Over-sampling Technique) method to oversample non-requirement data. It allowed us to achieve a balanced dataset by equalizing the representation of both non-requirement and requirement data. Hence, we trained each classifier (Step ③).

We used cross-validation to evaluate these models in Step ④. For this validation, we used the `Scikitlearn SearchGrid` function to tune our cross-validation parameter. After building the grid, we executed our `GridSearchCV` model for finding the best estimator parameter and the n-fold value. As a result, we employed a 10-fold cross-validation model to evaluate the performance of these classifiers. So, in each round of cross-validation, we trained our models using 80% of the corpus (2,494 sentences) and used the remaining 20% of the data (624 sentences) for testing the performance.

## V. EMPIRICAL RESULTS

In this section, we present the results of our benchmarking to answer our **RQ1** and **RQ2** in the designed case study. To evaluate the text classification performance, we collected subReddit posts to perform text classification and answer our research question. After that, we built the text classifier with machine learning by implementing five classifiers. We further added additional non-text features to enhance the performance of these techniques.

### A. Supervised learning using text features

We performed a comprehensive benchmark to find the best combination of text features and classifiers, as detailed in the previous section. In our case study, we manually labeled our subreddit student's sentences as "requirements" and "non-requirement" with the help of `Amazon Mechanical Turk` (see section III-B). In our research, we used both word embedding and sentence embedding methods to transform the textual data into a numerical format for training our classifiers.

TABLE I
RESULT OF MACHINE LEARNING CLASSIFIERS TRAINED ON WORD EMBEDDING MODELS FOR REDDIT DATA

| Model | TF-IDF | | |
|---|---|---|---|
| | Precision | Recall | F-Score |
| Logistic Regression | 0.63 | 0.57 | 0.59 |
| SVM | 0.65 | 0.58 | 0.61 |
| Naive Bayes | **0.67** | **0.65** | *0.65*\* |
| Random Forest | 0.60 | 0.55 | 0.57 |
| K-neighbour | 0.56 | 0.45 | 0.49 |
| Model | Word2vec | | |
| | Precision | Recall | F-Score |
| Logistic Regression | 0.60 | 0.55 | *0.57*\* |
| SVM | **0.61** | 0.53 | 0.56 |
| Naive Bayes | 0.59 | **0.58** | *0.58*\* |
| Random Forest | **0.60** | 0.53 | 0.56 |
| K-neighbour | 0.47 | 0.45 | 0.45 |

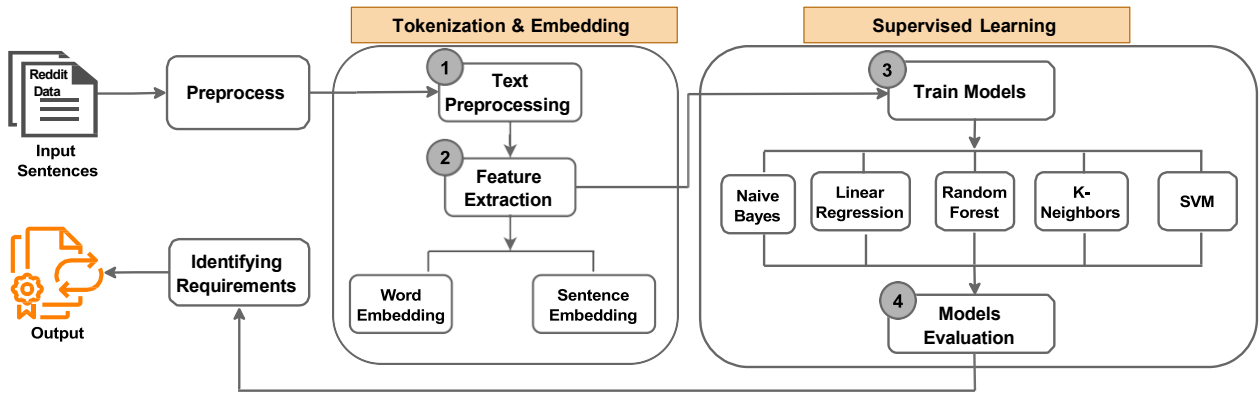

Fig. 5. Overview of the processes used for extracting the requirements from SubReddit data using NLP and text features.

TABLE II
RESULT OF MACHINE LEARNING CLASSIFIERS TRAINED ON SENTENCE EMBEDDING MODELS FOR REDDIT DATA

| MODEL | Sentence BERT | | | USE | | |
|---|---|---|---|---|---|---|
| Classifiers | Prec. | Recall | F-Score | Prec. | Recall | F-Score |
| Logistic Regression | 0.57 | 0.54 | 0.55 | 0.55 | 0.54 | 0.55 |
| SVM | 0.58 | 0.54 | 0.55 | 0.55 | 0.53 | 0.54 |
| Naive Bayes | 0.55 | 0.53 | 0.54 | 0.56 | 0.52 | 0.53 |
| Random Forest | 0.65 | 0.66 | **0.64*** | 0.61 | 0.64 | **0.62*** |
| K-neighbour | 0.42 | 0.44 | 0.43 | 0.44 | 0.45 | 0.44 |

TABLE III
THE RESULTS OF NAIVE BAYES CLASSIFIER WITH TF-IDF USING MULTIPLE FEATURES

| Features | Precision | Recall | F-Score |
|---|---|---|---|
| NB + TF- IDF | 0.67 | 0.65 | 0.65 |
| NB + TF- IDF + Interrogative | 0.82 | 0.76 | 0.78 |
| NB + TF- IDF + Keyword | 0.73 | 0.77 | 0.75 |
| NB + TF- IDF + Interrogative + Keyword | 0.84 | 0.75 | *0.79* |

Among the five classifiers, Naive Bayes showed the highest $F-score = 0.65$ when trained using TF-IDF. Similarly, when trained using Word2Vec word embedding, Naive Bayes showed the best F-score among all, with a score of 0.58. While the highest Precision was achieved by the Naive Bayes classifier trained with TF-IDF embedding. On the flip side, K-Neighbor is consistently performing worst among all the models. Table I shows the result of our benchmarking with word embedding. On the other hand, when we used Sentence-BERT, the Random Forest classifier showed the highest $F-score = 0.64$. For USE sentence embedding, Naive Bayes showed the best F-score among all classifiers, with a score of 0.62. Table II shows the result of our benchmarking with Sentence embedding. The overall use of Naive Bayes with TF-IDF resulted in a higher F-Score compared to other embedding methods.

*Naive Bayes model trained with TF-IDF demonstrated the highest level of accuracy in predicting and identifying requirements and achieving an F-score of 0.65.*

*B. Interrogation Tag and Requirements Keywords as Features*

In this section, we provide the results of our benchmarking analysis, which aims to address our **RQ2** in our chosen case study. In the previous section, we trained our five classifiers with word embeddings (TF-IDF and Word2Vec) and sentence embedding (Sentence-BERT and USE). We identified Naive Bayes with TF-IDF as the best-performing classifier in terms of F-score and in the context of our case study. However, all the classifiers showed a relatively weak overall performance. We decided to explore further if the use of additional labeling we introduced for sentences with requirement keywords (Section III-C) and for interrogative sentences (Section III-D) enhances the classifiers' performance. Hence, we used the Naive Bayes classifier using multiple attributes to identify requirements from SubReddit data shown in Table III.

Initially, we added the label whether the sentence is interrogative or not as a feature in addition to the text features when training a Naive Bayes classifier. Adding the annotation for interrogative sentences resulted in $F-score = 0.78$ for this classifier. We then used requirement keyword annotations as an additional feature, indicating whether the sentence includes a requirement keyword or not. There we obtained an $F-score = 0.75$. However, when we combined both interrogative sentence and keyword sentence annotations, we achieved the highest $F-score = 0.79$ to identify the requirement from SubReddit data.

*By including interrogative sentences and requirement keyword annotations as training features, the F-score of the Naive Bayes classifier improved to 0.79.*

VI. LIMITATIONS AND THREATS TO VALIDITY

There are several limitations to using natural language processing (NLP) and machine learning techniques for mining requirements from unstructured communication. We address limitations and potential threats to the validity of our research methods. We acknowledge that the organization of the threats to validity can be improved by categorizing them into specific classes of validity and providing steps to mitigate these threats.

*Construct Validity:* Construct validity is an important consideration in our study, particularly how requirements were defined and interpreted by crowd workers on Amazon Mechanical Turk. We have not provided a clear definition of the requirement, which may lead to a narrow interpretation based only on the examples given. This can introduce bias and limit staff understanding. Additionally, the absence of a criterion for accuracy in the labeling task raises concerns about potential bias due to compensatory behavior and noisy data. To address these concerns and improve construct validity, future iterations will provide a clearer definition of the requirement and include familiar statements to assess employee performance and reduce bias.

*Internal Validity:* The internal validity of our study is compromised by the potential bias of the labeling process performed through crowdsourcing using `Mturk`. This introduces the risk of faulty labels, which could introduce bias or inaccuracy in our results. To mitigate this threat, additional measures can be implemented, such as redundancy in labeling tasks and the employment of a survey-based approach to increase the reliability and accuracy of the labeling process.

*External Validity:* The external validity of our study is limited due to the narrow scope of our case study, we only consider the student posts from the subReddit forum based on specific keywords. While acting as a proof of concept, the case study is not generalized to all open forums and platforms. To enhance external validity, future research should encompass a broader range of online platforms and forums to ensure a more comprehensive understanding of user requirements during the pandemic.

*Conclusion Validity:* The conclusion validity of our study may be affected by two factors. Firstly, NLP and machine learning techniques depend on the availability of adequate and high-quality data. This can lead to inaccurate results. Also, the results analyzed may be limited by the complexity of unstructured communication. If the communications are too complex or use technical languages and are context specific. Hence, the classifiers might show different performances on other sources of communication. To address these concerns, future research can explore the use of domain-specific data and incorporate advanced preprocessing techniques to handle the complexities of communication more effectively. Lastly, The use of simple text representation and embedding also poses a limitation. Advances in NLP and deep learning can potentially lead to better models. We plan to use sequence-to-sequence models and bi-directional representation in the future. These advanced techniques have the capacity to capture more nuanced information from the text data, thereby improving the model's capability and strengthening the robustness of our conclusions.

## VII. VISION AND FUTURE WORK

In our study, we propose using communications related to a specific event, challenge, or topic as input to elicit software requirements. This approach serves as a complementary technique to traditional methods, aiming to bridge the requirements gap and streamline the software development process. Our initial investigation focused on employing machine learning techniques to assess the feasibility of mining requirements from these communications. The results demonstrated that conventional techniques can accurately identify sentences containing requirements. However, this initial step is just the beginning. Not all requirements extracted from communications are necessarily relevant or suitable for addressing through a software service. Consequently, the subsequent tasks of identifying and mapping the requirements to potential software services pose significant challenges. Furthermore, it is essential to evaluate the extent to which these requirements complement those elicited through traditional techniques. To achieve a comprehensive understanding, additional case studies need to be conducted to build upon our initial observations. In the future, our research aims to expand the dataset by incorporating more online platforms, allowing for a larger and more diverse collection of communications. Additionally, we plan to leverage advancements in machine learning techniques, particularly natural language processing (NLP) and large language models (LLMs), to perform more extensive analyses. By integrating LLMs into our research, we anticipate gaining deeper insights and achieving more accurate results in the elicitation of requirements from unstructured communications.

## VIII. CONCLUSION

We proposed a shift from mining requirements from software-related documents or feedback to mining communications related to a particular topic or event that the software is intended to support. We envision that mining such requirements can lead to a complementary set of requirements to the traditional elicitation techniques and eventually help in closing the requirements gap. We designed and conducted a case study to mine student requirements during the COVID-19 pandemic by mining their communication from Reddit. We used Amazon Mechanical Turk and manually labeled a set of 3,118 sentences as communicating a requirement or not. We then trained and evaluated five classification models where Naive Bayes showed the best performance. We then improved our models by adding interrogative sentences and requirements keywords in addition to the text embeddings. As a result, with an F-score of 0.79, we could detect the Requirement sentences from the Reddit posts. Our research has demonstrated that it is indeed feasible to identify sentences that contain requirements. This initial step lays the foundation for our proposed approach, which aims to map requirements from SubReddit communications related to a particular problem or scenario into software requirements.